\documentclass{elsart}
\usepackage{natbib}
\usepackage{epsfig,amssymb}

\begin{document}

\begin{frontmatter}



\title{Modelling the term structure of interest rates \'{a} la Heath-Jarrow-Morton but with non Gaussian fluctuations} 
\author[TCD]{Przemys\mbox{\l}aw Repetowicz}
\ead{repetowp@tcd.ie}
\ead[url]{www.maths.tcd.ie/$\sim$przemek}
\author[IIS]{Brian Lucey}
\ead{blucey@tcd.ie}
\author[TCD]{Peter Richmond}
\ead{richmond@tcd.ie}
\ead[url]{www.tcd.ie/Physics/People/Peter.Richmond}
\address[TCD]{Department of Physics, Trinity College Dublin 2, Ireland}
\address[IIS]{University of Dublin - School of Business Studies; University of Dublin - Institute for International Integration Studies (IIIS)}


\author{}

\address{}

\begin{abstract}
We consider a generalization of the Heath Jarrow Morton model
for the term structure of interest rates where
the forward rate is driven by Paretian fluctuations.
We derive a generalization of It\^{o}'s lemma
for the calculation of a differential of a Paretian
stochastic variable and use it to derive a Stochastic Differential
Equation for the discounted bond price.
We show that it is not possible to choose the parameters of the model
to ensure absence of drift of the discounted bond price.
Then we consider a Continuous Time Random Walk with jumps
driven by Paretian random variables and we derive the large time 
scaling limit of the jump probability distribution function (pdf).
We show that under certain conditions defined in text the large time scaling
limit of the jump pdf in the Fourier domain is 
$\tilde{\omega}_t(k,t) \sim \exp\left\{ -\mathfrak{K}/(\ln(k t))^2 \right\}$
and is different from the case of a random walk with Gaussian fluctuations.
\end{abstract}


\end{frontmatter}

\section{Introduction}
The Arbitrage Pricing Theory (APT) is a generalization of the Capital Asset Pricing Model 
(CAPM), and 
it states that capital assets 
are priced according to a set of factors. It does not say what 
these factors are, just that only idiosyncratic (firm 
specific) risk is priced. So, two different stocks could have two different 
APT formulations. 

However, the law of one price states 
that two identical assets should have
the same value on the financial market and
deviations from rational behavior of traders for example, 
significantly larger returns on Fridays the $13^{\mbox{th}}$, 
\citet{Lucey} have been evidenced.
Yet, market efficiency is an assumed prerequisite for CAPM to hold 
and contemporary finance still uses APT as a quantitative tool.
\citet{HJM} (see also \citet{MusielaRutkowski}) have formulated a general theory 
(to be referred to as the HJM model)
and a unifying framework 
for the existing APTs.

Presuming the forward interest rate and the bond price
to be a continuous time stochastic process
described by a Stochastic Differential Equation (SDE)
they derive the generic form of the SDE that ensures
that the bond price, related to the zero coupon bond 
(risk-less security issued by the central bank at 
a fixed interest rate called the spot rate), 
has a mean zero as a function of time.

The HJM model has been generalized by \citet{Sornette}, 
who, drawing on some ideas from string theory,
derives a general condition for a class of second order SDEs
that must be satisfied to ensure absence of arbitrage opportunities.
He obtains a whole class of, nonlinear second order SDEs
that, in some cases, can be solved exactly 
and that also provide extensions to the theory of pricing options 
\citet{MusielaRutkowski}.

These models have a rich structure but are based on Gaussian fluctuations.
In this paper we develop a corresponding theory
that allows the fluctuations to have an arbitrary distribution, 
in particular they may have power-law behavior at the high end.

\section{Theoretical analysis}
The purpose of this section is to provide overall understanding
of stochastic processes that involve asymptotically large, power-law distributed,
fluctuations. 
The analysis lacks mathematical rigor and draws more on 
physical ideas of Lang\'{e}vin processes 
(even though they are not explicitly mentioned in the text) 
than on the abstract theory of stochastic processes.
The derivation of strict mathematical foundations 
for the analysis (e.g. foundations for the {\bf Lemma} and the {\bf Statement} 
in the next section)
is left for future work.

\subsection{Definitions and heuristic reasoning}
The forward interest rate $f(t,T)$ is an interest rate 
that is agreed at time $t$ and that gives the trader a right to trade in an asset
at time $T$ (to be termed as maturity).
The underlying assumption in the HJM model is to describe
$f(t; T)$
through a following SDE:
\begin{equation}
f(t+\epsilon; T) = f(t; T) 
+ \int_t^{t+\epsilon} \alpha\left(\theta; T\right) d\theta 
+ \int_t^{t+\epsilon} \sigma\left(\theta; T\right) \stackrel{.}{\eta}(\theta)d\theta
\label{eq:DefIntRate}
\end{equation}
where $\epsilon > 0$.
Here the functions $\alpha$ and $\sigma$, called drift  
and volatility are in general dependent on the process itself,
and the fluctuations $\delta\eta = \stackrel{.}{\eta}(\theta)d\theta$
have a zero mean, are uncorrelated with one another for different times and
conform 
to a generic, non-Gaussian distribution with a density $\rho(\delta\eta)$.
An asset which at maturity has a value one Euro
has at some earlier time $t$ a smaller value that 
is expressed through a discount factor involving 
the forward rate (\ref{eq:DefIntRate}) as follows:
\begin{equation}
B\left(t; T\right) = B\left(t, f(t, T); T\right) 
=  \exp\left\{-\int_t^T f(t; \xi)\right\} d\xi
\label{eq:BondPrice}
\end{equation}
The bond price is related to the price of a zero-coupon bond $M(t)$
whose stochastic dynamics is given by a SDE that involves the spot rate of interest 
$f(\theta,\theta)$ and some fluctuations:
\begin{equation}
M(t) = \exp\left\{ - \int_0^t f(\theta,\theta) d\theta  + \int_0^t \stackrel{.}{\eta}(\theta)d\theta \right\}
\label{eq:discountfactor}
\end{equation}
and conditions are derived for the discounted bond price 
$M(t)B\left(t;T\right)$ to have a zero drift 
in order to ensure absence of arbitrage opportunities.
The zero-drift condition is also referred to as the ``martingale property'' 
and states that the conditional mean of the bond price at time $s > t$ 
discounted back to time zero equals the discounted bond price at time $t$.
Here $E\left[ M(s)B(s;T) \left| \mathcal{F}_t \right. \right] = M(t) B(t; T)$
and the conditioning is on the so called ``filtration'' $\mathcal{F}_t$ of information
meaning a set of information about the bond price up to time $t$   
(see  \citet{MusielaRutkowski}).

The main problem is now to derive 
a SDE for the bond price (\ref{eq:BondPrice}).
For computing a differential of a function of Gaussian fluctuations
one can use the lemma by \citet{Ito}. This lemma states that
the differential is obtained by expanding the function into 
a Taylor series to the first order in time and 
to the second order in the fluctuations and 
by replacing differentials of the fluctuations via their averages.
If the function of the stochastic process $f$ 
has a form given by equation (\ref{eq:BondPrice}) we may use a theorem by 
Girsanov [see \citet{MusielaRutkowski}] that simplifies the computation.

If, however, the fluctuations are not Gaussian, 
in particular if the likelihood for them 
to be large is much larger than in the  Gaussian case
then the higher moments and auto-covariances of the process may be infinite 
(as in the case of a L\'{e}vy or a Pareto distribution) 
and the procedure for computing the differential may fail.
It may also happen that the probability distribution 
of the squared fluctuation is of the same kind as that of the fluctuation itself 
which will certainly change the structure of the SDE that describes 
the bond price (\ref{eq:BondPrice}).  
In the following we examine the mathematical difficulties 
that arise. 
Since we are not aware of any generalization of It\^{o}'s
lemma that can be used in a straightforward manner for the calculation 
of the differential of the bond price, we must 
perform the computations from scratch.
The procedure consists in first working out a small change of the 
bond price, secondly in expressing it via the forward rate and
finally in  performing the limit $\epsilon \rightarrow 0$.
Here the change reads :
\begin{equation}
B(t+\epsilon;T) - B(t; T)
= B(t; T) 
\left[ 
e^{\int_t^{t+\epsilon} f(t+\epsilon,\xi) d\xi}
\exp\left(-\int_t^T \delta_\epsilon f(t;\xi) d\xi \right)
-1
\right]
\label{eq:Differential}
\end{equation}
where $\delta_\epsilon f(t;\xi) := f(t+\epsilon;\xi) - f(t;\xi)$.
Now we expand the right hand side in a Taylor series and retain 
terms proportional to $\epsilon$ neglecting higher powers of $\epsilon$. 
The expression in square brackets on the right hand side of 
(\ref{eq:Differential})
reads:
\begin{eqnarray}
\lefteqn{
\int_t^{t+\epsilon} f(t+\epsilon,\xi) d\xi 
+
\sum_{n=1}^\infty
\frac{(-1)^n}{n!}
\left(-\int_t^T \delta_\epsilon f(t;\xi) d\xi \right)^n =} \nonumber \\
&& 
\int_t^{t+\epsilon} f(t+\epsilon,\xi) d\xi 
-\int_t^T \int_t^{t+\epsilon} \alpha\left(\theta; \xi\right) d\theta d\xi 
-\int_t^T \int_t^{t+\epsilon} \sigma\left(\theta; \xi\right) 
\stackrel{.}{\eta}(\theta)d\theta d\xi \nonumber \\
&&+\sum_{n=2}^\infty \frac{(-1)^n}{n!}
\left( 
       \int_t^T \int_t^{t+\epsilon} 
       \sigma(\theta,\xi)\stackrel{.}{\eta}(\theta) d\theta d\xi
\right)^n 
\label{eq:Expansion}
\end{eqnarray}
Now the whole problem consists in a proper treatment of the products of fluctuations
$\prod_{i=1}^n \delta\eta(\theta_i)$ in the last term on the right hand side 
in (\ref{eq:Expansion}). This is the point where the analysis differs essentially
from the Gaussian case. We assume the fluctuations to conform to a L\'{e}vy
(or Paretian) distribution.
We set $\delta\eta = \mbox{L\'{e}vy}_\mu(0,1,0) dt$ which means that
$\stackrel{.}{\eta} = \delta\eta/dt$ 
is (parameters in brackets from the left to the right)
a Paretian distribution with a characteristic exponent
$\mu$, skewness zero, dispersion one and mean zero (to be termed standard L\'{e}vy variable).
The Paretian distribution has a pdf whose asymptotic expansion reads: 
\begin{equation}
\rho(\delta\eta) = \frac{1}{\pi} \sum_{n=1}^\infty \sin(\frac{\pi}{2} \mu n) \Gamma(\mu n + 1) \frac{dt^{n}}{x^{\mu n + 1}}
\label{eq:AsymptExp}
\end{equation}
From this expansion follows an important property for the product 
of many Paretian variables.

\noindent{\bf Lemma:}
\begin{enumerate}
\item[(A)] If $\delta\eta = \mbox{L\'{e}vy}_\mu(0,1,0) dt$ 
then $(\delta\eta)^n = \mbox{L\'{e}vy}_{\mu/n}(0,1,0) (dt/n)$ 
\item[(B)] For independent L\'{e}vy variables $\delta\eta_1$ and $\delta\eta_2$
we have 

$Z = \delta\eta_1 \delta\eta_2 \sim \mbox{L\'{e}vy}_\mu(0,1,0) (dt)^2$.
\end{enumerate}
Both statements regard the high end of the distribution.

\noindent{\bf Proof:}
\begin{enumerate}
\item[(A)] From the conservation of elementary probabilities
$f_Y(y)dy = f_X(x)dx$ it follows that
the pdf of the $n$-th power $Y = X^n$ of a Paretian variable 
$X = \mbox{L\'{e}vy}_\mu(0,1,0) dt$
reads:
\begin{eqnarray}
\lefteqn{
f_Y(y) = \frac{f_X(x)}{n x^{n-1}} = \frac{f_X(y^{1/n})}{n y^{1 - 1/n}}
       = \frac{1}{n y^{1-1/n}} \sum_{m=1}^\infty \frac{a_m dt^m}{(y^{1/n})^{\mu m + 1}}
       = \frac{1}{n} \sum_{m=1}^\infty \frac{a_m dt^m}{y^{(\mu/n) m + 1}}
}
\nonumber \\
       &&\Rightarrow Y \sim \frac{dt}{n} \mbox{L\'{e}vy}_{\mu/n}(0,1,0)
\end{eqnarray}
where $a_m$ denote coefficients in the expansion (\ref{eq:AsymptExp})
for the pdf of a Paretian variable.

\item[(B)] If $X_1,X_2= \mbox{L\'{e}vy}_\mu(0,1,0) dt$ are independent 
Paretian variables then the pdf of their product $Z = X_1 X_2$ reads:
\begin{eqnarray}
\lefteqn{
f_Z(z) = \int_1^z \int_1^z \delta( z - x_1x_2 ) f_X(x_1) f_X(x_2) dx_1 dx_2
}
\nonumber \\
&&= \int_1^z \frac{dx_1}{x_1} f_X(x_1) f_X(\frac{z}{x_1})
  = \int_1^z \frac{dx_1}{x_1} 
             \frac{(a_1 dt)}{\xi^{\mu+1}} \frac{(a_1 dt)}{(z/\xi)^{\mu+1}}
  = \frac{(a_1 dt)^2}{z^{\mu  +1}} \ln(z) \label{eq:ProdParetian} \\
&& \Rightarrow Z \sim (dt)^2 \mbox{L\'{e}vy}_{\mu}(0,1,0)
\end{eqnarray}
where (\ref{eq:ProdParetian})
holds only for large values of $z$ since we have replaced the pdfs 
by the first term of the asymptotic expansion (\ref{eq:AsymptExp}).
From the last expression in (\ref{eq:ProdParetian}) it follows that
,up to a logarithmic correction,
$Z$ is a Paretian with a squared dispersion  
{\bf q.e.d}.
\end{enumerate}
The {\bf Lemma} implies that the only factors proportional 
to the first power of $\epsilon$ in the last sum on the right hand side 
of equation (\ref{eq:Expansion}) are those which contain  
``diagonal'' terms of the kind $(\delta\eta(\theta_i))^n$.
Indeed from (B) it follows that the non-diagonal terms, ie those including products 
$(\delta\eta(\theta_1))^{n_1}(\delta\eta(\theta_2))^{n_2}\ldots (\delta\eta(\theta_p))^{n_p}$ 
of differentials of 
$\eta$ evaluated at different values of $\theta$ will all be proportional to 
higher powers of $\epsilon$ 
(since they are products of independent Paretian random variables)
and can, therefore, be neglected.
We emphasize the difference to the Gaussian case:

\fbox{
\begin{minipage}{\textwidth}
\noindent{\bf Statement}
The $n$-th power of a Gaussian with variance $dt$
is a random variable with variance $dt^n$.
The $n$th power of a Paretian with a dispersion $dt$ 
also has a dispersion $dt$.
\end{minipage}
}

We now obtain, after taking the limit $\epsilon \rightarrow 0$
 the following SDE for the bond price: 
\begin{eqnarray}
\frac{d_t B(t,T)}{B(t,T)} 
&=& f(t,t) - \int_t^T \alpha(t;\xi)d\xi +
\sum_{n=1}^\infty \frac{(-1)^n}{n!} \left( \int_t^T \sigma(t,\xi) d\xi \right)^n 
(\stackrel{.}{\eta}_\mu(t))^n \nonumber \\
&=& f(t,t) - \int_t^T \alpha(t;\xi)d\xi +
\exp\left[ - \left( \int_t^T \sigma(t,\xi) d\xi \right) \stackrel{.}{\eta}_\mu(t) \right]
- 1
\label{eq:LogDerivP}
\end{eqnarray}

In the next step we carry out the same procedure
for the discount factor $M(t)$ defined in (\ref{eq:discountfactor}).
Expanding $M(t+\epsilon) - M(t)$ in a Taylor series to arbitrary order
in the variable $\int_0^t \left( f(\theta,\theta) - \stackrel{.}{\eta}(\theta) d\theta \right) d\theta$,
as in  (\ref{eq:Expansion}),
we notice that after expanding the $n$-th (with $n>2$) power  of the integral
we get some terms that contain $f(\theta,\theta)$ and other terms that
contain only the fluctuations $\stackrel{.}{\eta}(\theta)$.
The former terms will be of order $\epsilon^2$ and may be neglected.
The only term left then is the $n$th power of the integral over fluctuations.
This we rewrite as a sum of increments $\delta\eta(t + i dt)$ 
at times $(t + i dt)$.
This corresponds to a division of nodes of the interval $[t, t+\epsilon]$
into $\epsilon/dt$ bins. Thus we have:
\begin{equation}
(\int_t^{t+\epsilon} \stackrel{.}{\eta}(\theta) d\theta)^n
=
\prod_{j=1}^n \left[ \sum_{i_j=1}^{\epsilon/dt} \delta\eta(t + i_j dt) \right]
=
\sum_{i=1}^{\epsilon/dt} ( \delta\eta(t + j dt) )^n + O(\epsilon^2)
= ( \stackrel{.}{\eta}(t) )^n \epsilon + O(\epsilon^2)
\end{equation}
Now notice that on the grounds of {\bf Lemma} and/or of the {\bf Statement}
that the power terms $( \delta\eta(t + j dt) )^n$ are $dt/n \mbox{L\'{e}vy}_ {\mu/n}(0,1,0)$
whereas all other terms 
$\delta\eta(t + j_1 dt)^{n_1} \cdot \ldots \cdot \delta\eta(t + j_p dt)^{n_p}$ 
that involve $p\ge 2$ different time values
are, since they are products of $p$ independent Paretian variables,
on the grounds of {\bf Lemma} (B) of order of $dt^p$.
Since their number is a polynomial of order $(\epsilon/dt)^p$ the total will 
be of order $\epsilon^p$ and can, therefore, be neglected.  

In the limit $\epsilon \rightarrow 0$ only the powers of Paretian variables survive
and we arrive at the following SDE
for the discount factor:
\begin{equation}
\frac{d_t M(t)}{M(t)} = -f(t,t) + \exp( \stackrel{.}{\eta}_\mu(t) ) - 1 
\label{eq:LogDerivM}
\end{equation}
Adding the logarithmic derivatives (\ref{eq:LogDerivM}) and (\ref{eq:LogDerivP})
we obtain the logarithmic derivative of the discounted bond price
$V(t) = M(t) B(t; T)$, ie the price at time zero of a bond that at maturity $T$
pays off 
one Euro and has, at some intermediate time $t$, the value $B(t; T)$.
\begin{equation}
\frac{d_t V(t)}{V(t)} =  
- \int_t^T \alpha(t;\xi)d\xi - 2 +
\exp\left[ - \left( \int_t^T \sigma(t,\xi) d\xi \right) \stackrel{.}{\eta}_\mu(t) \right] +
\exp( \stackrel{.}{\eta}_\mu(t) ) 
\label{eq:DiscountedBondPrice}
\end{equation}
This is an SDE for the discounted bond price that we must solve and then
require the Martingale condition 
$E\left[V(s) - V(t) \left| \mathcal{F}_t \right.\right] = 0$ 
for every $s > t$ to be satisfied in order to 
ensure non-arbitrage opportunities in the market with non Gaussian fluctuations.
Note that the time evolution of the logarithm of $V(t)$ is driven by two 
competing factors namely a negative ``deterministic'' drift 
corresponding to the first two terms on the right hand side in 
(\ref{eq:DiscountedBondPrice}) and a positive fluctuations
(the last two terms). Therefore the Martingale condition boils down
to requiring the average fluctuation to be equal to the deterministic drift.
Note also that, 
due to an exponential enhancement of the underlying fluctuations $\stackrel{.}{\eta}_\mu(t)$,
the fluctuations in (\ref{eq:DiscountedBondPrice}) are very large;
and huge when compared to Gaussian noise.
It is, therefore, not clear whether the absence of arbitrage opportunities can be
indeed ensured (see item (B) in the discussion section \ref{sec:Discussion}).

\subsection{Connections to other works in this subject matter \label{sec:Discussion}}
At this stage we make some remarks 
regarding the connection (A),(B) of our result to works by other authors 
and to an abstract representation (C) of infinitely divisible distributions. 
\begin{enumerate}
\item[(A)] The equations (\ref{eq:LogDerivP}) and (\ref{eq:DiscountedBondPrice}) 
cannot be directly related to the result (12) page 154 in \citet{Kleinert}.
This is because the author assumes the logarithmic characteristic function 
$\phi(k) := log E\left[ \exp(i k \stackrel{.}{\eta}(t)) \right]$ 
of the fluctuations $\stackrel{.}{\eta}(t)$ is 
an analytic function in $k$ 
or, in other words, that all moments of the fluctuations are finite
$\left< ( \stackrel{.}{\eta}(t) )^p \right> < \infty$,
(see expansion of the ``Hamiltonian'' -- equation (6) page 153 in \citet{Kleinert})
whereas we have $\phi(k) = log E\left[ i k \mbox{L\'{e}vy}_\mu(0,1,0) \right] = -|k|^\mu$ 
which is not analytic as $k\rightarrow 0$ 
(some of the constants $c_1,c_2,\ldots$ in expansion (6) page 153 may be infinite).  
\item[(B)] \citet{Bielecki}, using the framework of the abstract probability theory,
considers a Gaussian HJM model with several forward interest rates, 
so called default-able $g_i(t; T)$ $i=1,\ldots,K$ rates  and a default free
rate $f(t; T)$, subject to an ordering condition 
$g_K(t;t) > \ldots > g_1(t;t) > f(t;t)$ on the respective spot rates of interest.
They define a discounted Default-able Corporate Bond (DCB) $\widehat{Z}(t; T)$
that equals the consecutive discounted default-able bond prices $Z_i(t; T)$.
This is related to the default-able forward rates of interest $g_i(t;T)$,
and the discounted default free rate of interest $f(t; T)$ in time intervals 
$\tau_i < t < \tau_{i+1}$; intervals that are determined by the condition that the 
differences between the default-able bond prices $D_i(t; T)$ and the 
default free $B(t; T)$ price do not exceed a given threshold. 
In other words they consider a process of ``credit migrations'' 
where the DCB is taken as $D_k(t; T)$ at time inception
and is subsequently replaced by $D_i(t; T)$ with $i<K$ once 
differences benchmarked on the default free price $B(t; T)$
exceed a given threshold (see Fig.\ref{fig:DCBEvolution} ).
In addition, certain rates $\delta_i \in [0,1]$ of recovery 
(returns of losses from the Treasury Bond -- central bank)
are assumed in each time interval $t \in [\tau_i,\tau_{i+1}]$.
Finally conditions on the absence of arbitrage opportunities  for the DCB
are deduced.

\begin{figure}[!h]
\centerline{\psfig{figure=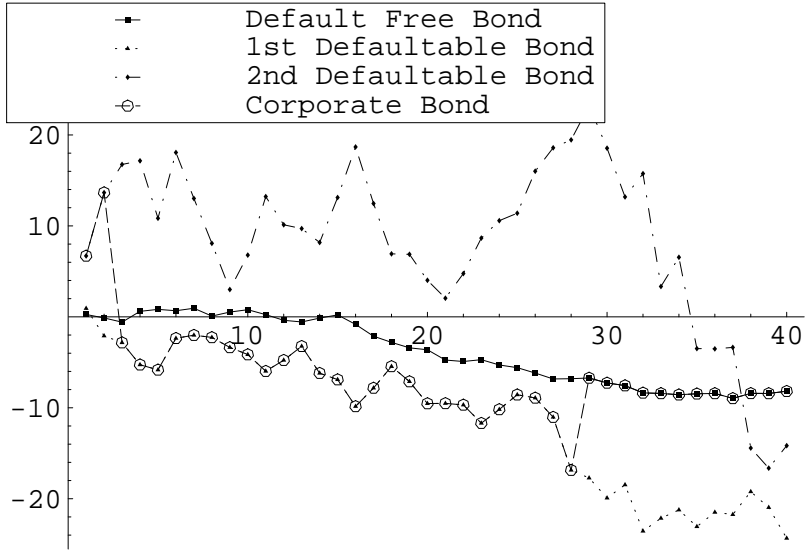,width=0.8\textwidth,angle=0}}
\caption{Time evolution of the logarithm of price of the 
         Discounted Corporate Bond as considered in \citet{Bielecki}.
         The evolution starts with the most highly default-able bond (Diamonds)
         and goes down to the less default-able (Triangles)
         and default free (Boxes) bonds once a threshold difference
         from the default free logarithm of price is reached.
         \label{fig:DCBEvolution}}
\end{figure}

\noindent This is certainly a very sophisticated ``real world'' model
yet it does not compare to our approach because 
in the presence of large Paretian fluctuations (see {\bf Statement}), 
that are presumed in our approach, absence of arbitrage opportunities 
cannot be ensured.
Indeed, a closer explanation of the time evolution
(\ref{eq:DiscountedBondPrice}) of the discounted bond price $V(t)$
ensures that it is impossible to choose the drift $\alpha(t; T)$
and the volatility $\sigma(t; T)$ such that the discounted bond price 
as a whole has a zero drift, ie the mean of a small increment 
$V(s) - V(t)$ for $s>t$ of the discounted bond price conditioned 
on the past history of the price process equals zero under 
a certain probability measure.
This follows from a following 
 ``waving hand argument'' .
Due to
an exponential enhancement of, already very strong, Paretian fluctuations;
the mean of the right hand side of 
equation (\ref{eq:DiscountedBondPrice})  is infinite. 
Therefore the positive ``stochastic jumps'' dictated by the last two terms 
in equation (\ref{eq:DiscountedBondPrice}) are, 
for sufficiently large times $t$,  arbitrarily large and cannot be offseted 
by any finite value of the negative ``deterministic'' drift coefficient 
$- \int_t^T \alpha(t;\xi)d\xi - 2$.
Thus the discounted bond price drifts
toward positive values like a biased random walk.
Since
it is impossible to damp the biased movement of the discounted bond price $V(t)$.
Further theoretical analysis
of the model will be pursued in a different direction namely we shall
work out a distribution of random times needed
for the biased movement to reach a certain threshold $\mathfrak{x}$ value of the log-price,
known as first hitting times 
\begin{equation}
T_\mathfrak{x} = \mbox{min}\left[ t \left| \ln(V(t) = \mathfrak{x} \right. \right]
\label{eq:1stHittingTimeDef}
\end{equation}
of the price.
This can be done by writing down a Fokker-Planck differential equation,
an equation that may include fractional derivatives \citet{SamkKilb,Meerschaert}
and that describes the spatio-temporal evolution of the probability
for the discounted bond price to have a given value at a given time 
(to be termed occupation probability).
The equation may be solved in the asymptotic limit of large times and 
the probability distribution of the first hitting times 
will  be derived [\citet{Redner}] from it.
This approach will be pursued in section (\ref{sec:1stHittingTimes}).

\item[(C)] Since the stochastic process (\ref{eq:DefIntRate}) 
that describes the forward rate of interest 
is an Infinitely Divisible Distribution (IID)
it must have a representation in terms of the L\'{e}vy-Khintchin 
formula (e.g. Theorem 3.1.11 page 41 in \citet{Meerschaert})
for the log-characteristic function of an IID.
Indeed the stochastic dynamics (\ref{eq:DefIntRate}) of the forward rate
can be considered as a 
a compound Poisson process
$f(t; T) = \alpha(t; T) + \sigma(t; T)\sum_{j=1}^{N_t} \delta\eta_j$ with $N_t = \mbox{Poisson($c_\epsilon$)}$
and $\delta\eta_j = \mbox{L\'{e}vy}_\mu(0,1,0)$
with a very large rate $c_\epsilon$. 
This corresponds to taking a small positive $\epsilon >0$ and
a measure $\phi(dx)$ in the L\'{e}vy-Khintchin formula equal to 
$\phi(dx) \sim 1/x^{\alpha + 1}$. In this case
$c_\epsilon = \int_\epsilon^\infty dx/x^{\alpha + 1}$ can be arbitrarily large
for $\epsilon$ sufficiently small. Thus the jump pdf 
$\rho(\delta\eta) \sim c_\epsilon^{-1} 1/(\delta\eta)^{\alpha + 1}$ 
is related to a Paretian distribution for large values of the jump size $\delta\eta$.
\end{enumerate}

\section{Occupational probability of the discounted bond price\label{sec:1stHittingTimes}}
Now we formulate an equation
for the occupational probability 
$\mathcal{P}\left(x,t\right) = P\left( \ln V(t) = x \right)$ 
for the logarithm $\ln V(t)$ of the discounted bond price in (\ref{eq:DiscountedBondPrice})
to attain a value $x$ at time $t$.
Note that since the process $\ln V(t)$ is a Continuous Time Random Walk (CTRW) with jumps occurring
at random times
the probability density function (pdf) $\phi_n(x,t)$ of reaching $x$ at time $t$ after having performed $n$ steps
is a convolution of the pdf $\phi_{n-1}\left(x',t'\right)$ and a jump probability 
$\omega_t\left(x - x', t - t'\right)$. Here
\begin{equation}
\phi_n\left(x,t\right) = 
\int_{\-\infty}^\infty \int_0^t
\omega_t\left(x - x', t - t'\right) \phi_{n-1} \left(x',t'\right) dt' dx'
\end{equation}
Now we use the fact that the occupational probability $\mathcal{P}\left(x,t\right)$
is a convolution in time of the probability density of reaching $x$ at time $t'$
after an arbitrary number of steps $n$ and the probability density of being
unmoved in the time interval $\left[t', t\right]$. 
We obtain [\citet{Montroll,Balescu}]
the following formula for the Fourier-Laplace transform 
$\tilde{P}(k, s) := \mathcal{F}_x \mathcal{L}_t\left[ \mathcal{P}\left(x,t\right) \right](k, s)$
(with respect to $x$ and to time $t$ respectively)  of the occupational probability.
\begin{equation}
\tilde{P}(k, s) = \frac{1 - \tilde{\phi}(s)}{s} \frac{1}{1 - \tilde{\omega}_t(k,s)}
\label{eq:OccupProb}
\end{equation}
where 
$\tilde{\omega}_t(k,s) = \mathcal{F}_x \mathcal{L}_t\left[\omega_t(x,t)\right] = \int_{-\infty}^\infty \int_0^\infty \exp(\imath k x - s t) \omega_t(x,t) dx dt$ is the Fourier-Laplace transform of the jump probability density
and $\tilde{\phi}(s) = \tilde{\omega}_t(0,s)$ is the Laplace transform of the waiting time
pdf. Thus in order to find the occupational probability we 
need to work out the Fourier-Laplace transform of the jump pdf. 

Denoting $\Sigma := \int_t^T \sigma(t,\xi) d\xi$ and 
$\Delta := \int_t^T \alpha(t;\xi)d\xi$
we express the jump probability as:
\begin{eqnarray}
\lefteqn{
\omega_t(x-x', t-t') = P\left( \ln V(t) - \ln V(t') = x - x' \right)}  \label{eq:JumpPDFI} \\
&&= P\left( \int_{t'}^t \left(
                         \exp\left[ - \Sigma \stackrel{.}{\eta}_\mu(\xi) \right] +
                         \exp( \stackrel{.}{\eta}_\mu(\xi) ) 
                        \right) d\xi 
            = x - x' + \int_{t'}^t (\Delta(\xi) + 2)d\xi \right) \label{eq:JumpPDFII} \\
&&= P\left( 
           \frac{\Delta t}{N} 
            \sum_{j=1}^N \left( 
                         \exp\left[ - \Sigma \zeta_j \right] +
                         \exp( \zeta_j ) 
                        \right) 
            = \left( x - x' + \int_{t'}^t (\Delta(\xi) + 2)d\xi\right)
     \right) \label{eq:JumpPDFIII} \\
&&= \int\limits_{-\infty}^\infty\ldots\int\limits_{-\infty}^\infty 
    \delta\left(
           \frac{\Delta t}{N}
           \sum_{j=1}^N \left( 
                         e^{ - \Sigma \zeta_j } +
                         e^{ \zeta_j } 
                        \right) 
            - \left( x - x' + \int_{t'}^t (\Delta(\xi) + 2)d\xi\right)
          \right)
    \left[ \prod_{j=1}^N \rho(\zeta_j) d\zeta_j \right]
\label{eq:JumpPDFIV}
\end{eqnarray}
where the equality (\ref{eq:JumpPDFI})=(\ref{eq:JumpPDFII}) follows from
the SDE (\ref{eq:DiscountedBondPrice}). The next equality
follows from the division of the time interval $[t',t]$ into $N$
intervals of length $\Delta t/N$ and from the definition of a Riemann integral.
The final equality (\ref{eq:JumpPDFIII})=(\ref{eq:JumpPDFIV}),
in which $\delta$ means the Dirac delta function,
follows from the fact that the values of fluctuations $\zeta_j = \stackrel{.}{\eta_\mu}(\xi_j)$ 
at intermediate
times $\xi_j = t' + j (t - t')/N$ are iid standard L\'{e}vy variables.
Performing the Fourier transform of the jump pdf with respect to $x-x'$ 
results in replacing the delta function in the integrand in (\ref{eq:JumpPDFIV})
by an exponential $\exp\left\{\imath k \ldots\right\}$ 
and produces a following path integral
\begin{equation}
\tilde{\omega}_t(k,t-t') =
\frac{1}{2\pi}
\int\limits_{-\infty}^\infty\ldots\int\limits_{-\infty}^\infty 
\exp\left\{\imath k 
\int_{t'}^t \mathcal{H}(\xi,\stackrel{.}{\eta}_\mu(\xi))
            d\xi
\right\}
D\left[ \vec{\stackrel{.}{\eta}_\mu} \right]
\label{eq:PathIntegral}
\end{equation}
where the ``Hamiltonian'' $\mathcal{H}(\xi,\stackrel{.}{\eta}_\mu(\xi))$ reads: 
\begin{eqnarray}
\mathcal{H}(\xi,\stackrel{.}{\eta}_\mu(\xi)) 
&=& \mathcal{H}_s(\stackrel{.}{\eta}_\mu(\xi)) - \mathcal{H}_d(\xi)  \\
&=&
             \left(
             \exp\left[ - \Sigma \stackrel{.}{\eta}_\mu(\xi) \right] +
             \exp( \stackrel{.}{\eta}_\mu(\xi) ) 
            \right)
            -
            \left(
             \Delta(\xi) + 2
            \right) 
\label{eq:Hamiltonian}
\end{eqnarray}
Here we divided the ``Hamiltonian'' into two parts 
the ``stochastic'' part \\ $\mathcal{H}_s(\stackrel{.}{\eta}_\mu(\xi))$ 
that depends on the fluctuations $\stackrel{.}{\eta}_\mu(\xi)$ and 
the ``deterministic'' part $\mathcal{H}_d(\xi)$ that depends on the drift function
$\Delta(\xi)$.
The measure
$D\left[ \vec{\stackrel{.}{\eta}_\mu} \right]$
is expressed via pdfs of  standard L\'{e}vy variables 
$\stackrel{.}{\eta}_\mu(\xi_j)$
as follows:
$D\left[ \vec{\stackrel{.}{\eta}_\mu} \right] = 
\mbox{lim}_{N\rightarrow \infty} 
\left[ \prod_{j=1}^N \rho(\stackrel{.}{\eta}_\mu(\xi_j)) d\stackrel{.}{\eta}_\mu(\xi_j) \right]$
Note that since the ``Hamiltonian'' depends only on the fluctuations
$\stackrel{.}{\eta}_\mu(\xi)$ and not on their time derivative
$\stackrel{..}{\eta}_\mu(\xi)$ 
\footnote{This would correspond to a propagator of the Schr\"{o}dinger equation in quantum mechanics
with a Hamiltonian that only depends on the coordinates of the particle and not on its velocities.}
the path integral (\ref{eq:PathIntegral}) is expressed via ordinary 
Riemann integrals only.
Here:
\begin{equation}
 \tilde{\omega}_t(k,t-t') =
\frac{1}{2\pi}
e^{ -\imath k \int_{t'}^t \mathcal{H}_d(\xi) d\xi }
\cdot
\mbox{lim$_{N\rightarrow \infty}$}
\left[
\mathfrak{f}\left(\frac{k(t - t')}{\varpi(N)}\right)
\right]^N
\label{eq:JumpPdf}
\end{equation}
where
\begin{equation}
\mathfrak{f}(k)  := 
\int\limits_{-\infty}^\infty \exp\left\{\imath k  \mathcal{H}_s(\zeta) \right\}
\rho(\zeta) d\zeta
= \int\limits_{-\infty}^\infty \exp\left\{
\imath k(e^\zeta + e^{-\Sigma \zeta})\right\} \rho(\zeta) d\zeta
\label{eq:PathIntLaplTrDef}
\end{equation}
and $\varpi(N)$ is a number of subintervals into which we split the interval $[t',t]$
so that the the path integral does not diverge when $N \rightarrow \infty$.

In order to get additional  insight into the solution of the problem
in our further calculations we replace the integrand in the definition
of $\mathfrak{f}(k)$ by a sum of two terms in which one of the 
$e^{(\ldots\zeta)}$ terms is neglected. The applicability of this approximation
is yet to be examined. 

\noindent{\bf Approximation:}

\begin{equation}
\mathfrak{f}(k) \simeq \frac{1}{2}
\left(
\mathfrak{F}(1,k) + \mathfrak{F}(-\Sigma, k)
\right)
\label{eq:Approx1}
\end{equation}

where 
\begin{equation}
\mathfrak{F}(a,k)
:= \int\limits_{-\infty}^\infty \exp\left\{\imath k e^{a\zeta}\right\} \rho(\zeta) d\zeta =
\frac{-1}{2\pi a} \int_{-\infty}^\infty \exp(-|\xi|^\mu) (\frac{k}{\imath})^{\imath \xi/a} \Gamma(-\imath \xi/a) d\xi
\label{eq:Approx2}
\end{equation}
In the second equality on the right hand side in (\ref{eq:Approx2})
we have substituted for $e^{\zeta}$ and 
used the integral representation \\
$\rho(\xi) = 1/(2\pi) \int_{-\infty}^\infty \exp(-\imath \xi - \left|\xi\right|^\mu) d\xi$
of the L\'{e}vy distribution.  
We then have changed integration limits in one of the integrals by making use of the Cauchy theorem
from complex analysis and 
replaced one of the integrals by the Gamma function making use of its integral representation. 
The minus sign in the last expression in (\ref{eq:Approx2}) stems from changing the integration
limits after applying the Cauchy theorem.
In further calculations we exploit the following results:

\noindent{\bf Laplace Transforms:} 

\begin{equation}
\mathcal{L}_t\left[t^\alpha\right](s) = \Gamma(\alpha+1)/s^{\alpha+1}
\label{eq:mathfrakA}
\end{equation}

\begin{equation}
\mathcal{L}_t\left[e^{\imath k t}\right](s) = 1/(s-\imath k)
\label{eq:mathfrakB}
\end{equation}

\begin{equation}
\mathcal{L}_t\left[f(k t)\right](s) = 1/k \mathcal{L}_t\left[f(t)\right](s/k)
\label{eq:mathfrakC}
\end{equation}

\begin{equation}
\mathcal{L}_t\left[\frac{1}{(\ln(t))^n}\right](s) = 
\frac{(-1)^n}{s (\ln(s))^n} \sum_{p=0}^\infty (-1)^p \frac{\mathfrak{k}_p}{(\ln(s))^p}
\label{eq:mathfrakCa}
\end{equation}

\begin{equation}
\mathcal{L}_t\left[\frac{e^{-\alpha t}}{(\ln(k t))^n}\right](s) 
= \frac{1}{(\alpha + s)} \frac{1}{(\ln(\frac{k}{s}))^n}
\sum_{p=0}^\infty
\frac{\mathcal{C}_p}{(\ln(\frac{k}{s}))^p}
\label{eq:mathfrakCb}
\end{equation}

\noindent{\bf The Gamma Function} 

\begin{equation}\Gamma(x) \Gamma(1 - x) = \pi / \sin(\pi x) \quad\mbox{for}\quad x \in \mathbb{C}
\label{eq:mathfrakD}
\end{equation}

\noindent{\bf Convolutions of Laplace transforms}

\begin{eqnarray} 
\lefteqn{
\left( \frac{1}{s} \otimes \frac{1}{s} \right) = \frac{1}{s}} \label{eq:mathfrakE1} \\
\lefteqn{
\left( s^n \otimes \frac{1}{s (\ln(s))^m} \right) = 
\frac{s^n}{(\ln(s))^m} \sum_{p=0}^\infty \mathfrak{O}_{m,p} \frac{\mathfrak{m}_p}{(\ln(s))^p}
\;\mbox{,}\;
}
\label{eq:mathfrakE3} \\
\lefteqn{\left( \frac{1}{s (\ln(s))^n} \otimes \frac{1}{s (\ln(s))^m} \right) 
= \frac{1}{s(\ln(s))^{n+m}}
\sum_{p=0}^\infty   \mathfrak{O}_{m,p} \frac{\mathfrak{n}_p}{(\ln(s))^p}
\;\mbox{,}\;
}
\label{eq:mathfrakE4}
\end{eqnarray}
where $\mathfrak{m}_p := \sum_{q=1}^n  (-1)^q/q^{p+1}$,
\mbox{$\mathfrak{n}_p := \sum_{q=1}^\infty  \mathfrak{l}_q(n)/q^{p+1}$}
and \mbox{$\mathfrak{O}_{m,p} := (m+p-1)!/(m-1)!$}.
Here $\otimes$ denotes a convolution \\
$(f(s) \otimes g(s))(s) := \int_0^s f(\xi) g(s - \xi) d\xi$ and 
the remaining parameters are defined 
and the results are derived in Appendix.

Now we work out the Laplace transform of the function 
$\mathfrak{f}(t)$. We get:
\begin{equation}
\tilde{\mathfrak{f}}(s) = \mathcal{L}_{t}\left[ \mathfrak{f}(t) \right](s) =
\frac{1}{2}\left(\tilde{\mathfrak{F}}(1,s) + \tilde{\mathfrak{F}}(-\Sigma, s)\right)
\label{eq:PartOfKernelDefI}
\end{equation}
where
\begin{eqnarray}
\lefteqn{
\tilde{\mathfrak{F}}(a,s) = \frac{-1}{2\pi a}
\int\limits_{-\infty}^\infty \exp\left(-|\xi|^\mu + \pi/(2 a) \xi\right)
\frac{1}{s^{1 + \imath \xi /a}} \frac{\imath \pi}{\sinh(\pi \xi/a)} d\xi} 
\label{eq:PartOfKernelDefIIa} \\
&&= \frac{1}{2 a s}
\int\limits_{-\infty}^\infty \exp\left(-|\xi|^\mu + \pi/(2 a) \xi\right)
\frac{\sin(\xi/a \ln(s))}{\sinh(\pi \xi/a)} d\xi 
\label{eq:PartOfKernelDefIIb} \\
&&= \frac{1}{2 a s}
\int\limits_{-\infty}^\infty 
\exp\left(-|\xi|^\mu + \pi/(2 a) \xi\right)
\cdot
\frac{\xi}{\sinh(\pi \xi/a)}
\cdot
\frac{\sin(\xi (\ln(s)/a))}{\xi} d\xi
\label{eq:PartOfKernelDefIIc}
\end{eqnarray}
In the first equality (\ref{eq:PartOfKernelDefIIa}) we make use of  (\ref{eq:mathfrakA})
and of the result (\ref{eq:mathfrakD}) concerning a product of Gamma functions.
In (\ref{eq:PartOfKernelDefIIb}) we write the term $s^{\imath \xi/a}$ as 
$\exp(\imath \xi/a \ln(s))$ and we disregard the imaginary part owing 
to the fact that the result is real. In  (\ref{eq:PartOfKernelDefIIc})
we separate the integrand into a product of terms that will be handled later on. 

Now we find a small-$s$ approximation to the function
$\tilde{\mathfrak{F}}(a,s)$. 
Since for small positive $s$ the last part of the integrand in 
(\ref{eq:PartOfKernelDefIIc}) is a very ``high frequency'' periodic function of $\xi$
with a maximum $(\ln(s)/a)$ at $\xi = 0$ and a period of length $2\pi a/|\ln(s)|$
we apply the following approximations (to be termed steepest descent approximations)
for our calculations. 
\begin{itemize}
\item[$\mathfrak{b}(1)$] Expand the first term 
$\mathfrak{I}(\xi) := \exp\left(-|\xi|^\mu + \pi/(2 a) \xi\right)$
in the integrand 
(\ref{eq:PartOfKernelDefIIc}) in a Taylor series in $\xi$ to the second order around zero.
Note that the maximum of the whole integrand (the product of all three terms in 
(\ref{eq:PartOfKernelDefIIc})) is bigger than zero but, since 
the expression $\sin(\xi (\ln(s)/a))/\xi$ behaves for small positive $s$
as $\delta(\xi/a)$, the maximum of the whole integrand will be arbitrarily
close to zero for sufficiently small $s$.

\item[$\mathfrak{b}(2)$] Confine the integration range 
to the first period $\left[-\pi a/|\ln(s)|,\pi a/|\ln(s)|\right]$ 
of the ``high-frequency'' periodic function in the last term of the integrand.
\item[$\mathfrak{b}(3)$] Replace the middle and the last terms in the integrand by 
a small $\xi$ expansion around $\xi=0$.
\end{itemize}

This yields a following result:
\begin{eqnarray}
\tilde{\mathfrak{F}}(a,s) &\simeq&  \frac{1}{2 a s}
\int\limits_{-\pi a/|\ln(s)|}^{\pi a/|\ln(s)|}
\left( \mathfrak{I}(0)  + \frac{1}{2} \mathfrak{I}^{''}(0_+) \xi^2 \right)
\cdot
\frac{a}{\pi}
\cdot
\frac{\ln(s)}{a} d\xi \nonumber \\
&=& \frac{1}{2 a s} 
\cdot
\frac{\ln(s)}{\pi}
\cdot
\left[
\mathfrak{I}(0) \frac{2\pi a}{|\ln(s)|} + 
\frac{1}{3} \mathfrak{I}^{''}(0_+)
\left(\frac{\pi a}{|\ln(s)|}\right)^3
\right] \nonumber \\
&=& 
 \frac{1}{s}
-
\frac{\pi^2}{6} |\mathfrak{I}^{''}(0_{+})| a^2 \frac{1}{s \ln(s)^2}
+
\frac{1}{s} O\left( \frac{1}{\ln(s)^4} \right)
\label{eq:PartOfKernelDefApprox}
\end{eqnarray}

This calculation is incomplete in that we did not explain sufficiently the nature of 
approximation used by computation of the integral. In fact we should have expanded
the integrand around its maximum $\xi_0=\xi_0(s)$, calculated the integral and only then
considered the result in the limit of small positive $s$. This
would only change the value of the parameter $\mathfrak{I}^{''}(0_{+})$ in the result.
In our calculations this parameter is left unknown. 

Now we invert the Laplace transform $\tilde{\mathfrak{f}}(s)$ 
and, on the grounds of  (\ref{eq:mathfrakCa}) and of (\ref{eq:PartOfKernelDefI}),
we get:
\begin{equation}
\tilde{\mathfrak{f}}(t) = \mathcal{L}_s^{-1}\left[\tilde{\mathfrak{f}}(s)\right](t) = 
1 - \frac{\mathfrak{K}}{(\ln(t))^2} + O\left(\frac{1}{(\ln(t))^4}\right)
\label{eq:fFctTimeDomain}
\end{equation}
where $\mathfrak{K}$ is some parameter that is unknown.
Inserting (\ref{eq:fFctTimeDomain}) into 
 (\ref{eq:JumpPdf})
we obtain the jump probability in the limit of large times:
\begin{eqnarray}
 \tilde{\omega}_t(k,t) &=&
\frac{1}{2\pi}
e^{ -\imath k \int_{0}^t \mathcal{H}_d(\xi) d\xi }
\cdot
\mbox{lim$_{N\rightarrow \infty}$}
\left[ 1 - \frac{\mathfrak{K}}{(\ln(\frac{k t}{\varpi(N)}))^2} 
\right]^N
\label{eq:FinalResultJumpPdf} \\
&=& \frac{1}{2\pi}
e^{ -\imath k \int_{0}^t \mathcal{H}_d(\xi) d\xi }  
\cdot
\mbox{lim$_{N\rightarrow \infty}$}
\exp\left\{ -\frac{N\mathfrak{K}}{(\ln(\frac{k t}{\varpi(N)}))^2}\right\}
\label{eq:FinalResultJumpPdfa} \\
&=& \frac{1}{2\pi}
e^{ -\imath k \int_{0}^t ( \Delta(\xi) + 2 ) d\xi }  
\cdot
\exp\left\{ -\frac{\mathfrak{K}}{(\ln(k t))^2}\right\}
\label{eq:FinalResultJumpPdfb}
\end{eqnarray}
where in (\ref{eq:FinalResultJumpPdfb}) we have chosen
the number of subintervals $\varpi(N)$ as 
\mbox{$\varpi(N) = (k t)^{\sqrt{N}+1}$}
to ensure that the limit is finite.
Now we use result (\ref{eq:mathfrakCb}) to calculate the Fourier-Laplace
transform $\tilde{\omega}_t(k,s)$ of the jump pdf and to obtain
the Fourier-Laplace transform of the occupational probability 
$\tilde{P}(k, s)$ of the discounted bond price from
(\ref{eq:OccupProb}).
We assume for simplicity that the function $\Delta(\xi)$ does not depend on time.
\begin{eqnarray}
\lefteqn{
\tilde{\omega}_t(k,s) = 
\mathcal{L}_t\left[e^{ - \imath k \mathcal{D} t}
\cdot
\exp\left\{ -\frac{\mathfrak{K}}{(\ln(k t))^2}\right\}
\right](s)
} \label{eq:JumpPDFFourLapl1}\\
&&=\sum_{n=0}^\infty \frac{(-\mathfrak{K})^n}{n!} 
\mathcal{L}_t\left[\frac{e^{-\imath k \mathcal{D} t}}{(\ln(k t))^{2 n}}\right](s) 
\label{eq:JumpPDFFourLapl2}\\
&&=\frac{1}{(\imath k \mathcal{D} + s)}
\left(
1  - \mathfrak{K} \frac{1}{(\ln(\frac{k}{s}))^2}
+ O\left(  \frac{1}{(\ln(\frac{k}{s}))^3} \right)
\right)
\label{eq:JumpPDFFourLapl3}
\end{eqnarray}
where 
$\mathcal{D} := ( \Delta + 2 ) $.
Inserting $k=0$ in (\ref{eq:JumpPDFFourLapl3}) we get 
$\tilde{\phi}(s) = \tilde{\omega}_t(0,s) = 1/s$
what implies that the waiting time pdf $\phi(t)$
is constant $\phi(t) = 1$ as a function of time and is not 
normalisable $\int_0^\infty \phi(t) dt = \infty$!
Because of this constructing the occupational probability
$\mathcal{P}\left(x,t\right)$
from  (\ref{eq:OccupProb})
in a process where the transition of the walker to $x$ in time $t$
consists of performing a certain number of jumps and 
of being immobile for some time may be impossible.
This is because the Fourier Laplace transform $\tilde{\omega}_t(k,s)$
may  be bigger than one and thus 
the formula (\ref{eq:OccupProb}) cannot be applied.

Note that if the drift $\Delta(\xi)$ satisfies a condition:
$\Delta + 2 = 0$ 
then the Fourier transform of the jump pdf 
decays 
as $\tilde{\omega}_t(k,t) \sim \exp\left\{ -\mathfrak{K}/(\ln(k t))^2 \right\}$
This decay is much slower than the inverse of any positive power of $t$.
The interpretation of this result and fitting the model to high-frequency
financial data is left for future work.

\section{Master equation for the discounted bond price}
In this section we derive a partial differential equation that is satisfied by
the jump pdf $\omega_t\left(x, t\right)$. The derivation bases on the representation
(\ref{eq:PathIntegral}) of the jump pdf as a path integral.
Note that equation (\ref{eq:JumpPdf}) can be written as follows:
\begin{equation}
 \tilde{\omega}_t(k,t-t') =
\frac{1}{2\pi}
\exp\left[ -\imath k \int_{t'}^t \left( \Delta(\xi) + 2 \right) d\xi \right]
\cdot
\exp\left[ - \int_{t'}^t \tilde{\mathfrak{H}}^{(\Sigma(\xi))}(k) d\xi \right]
\label{eq:MasterEqJumpPdf}
\end{equation}
where 
\begin{equation}
\tilde{\mathfrak{H}}^{(\Sigma)}(k) := -\ln(\mathfrak{f}(k))
= \ln\left[\int\limits_{-\infty}^\infty \exp\left\{
\imath k(e^\zeta + e^{-\Sigma \zeta})\right\} \rho(\zeta) d\zeta\right]
\end{equation}
The function $\tilde{\mathfrak{H}}^{(\Sigma)}(k)$ 
can be viewed as a ``log-characteristic function''
related to a certain integral transformation  of the jump pdf
$\rho(\zeta)$, a transformation defined by equation (\ref{eq:PathIntLaplTrDef}).
If we now assume $\Delta(\xi)$ and $\Sigma(\xi)$ to be constant as functions of $\xi$
then from (\ref{eq:MasterEqJumpPdf}) we obtain an equation 
for the Fourier transform of the jump pdf
\begin{equation}
\partial_t \tilde{\omega}_t(k,t) = 
\left( \Delta + 2 \right) (-\imath k ) \tilde{\omega}_t(k,t) 
- \tilde{\mathfrak{H}}^{(\Sigma)}(k) \tilde{\omega}_t(k,t) 
\end{equation}
and, after taking the inverse Fourier transform,
an equation for the jump pdf itself:
\begin{equation}
\partial_t \omega_t(x,t) = 
-\left( \Delta + 2 \right) \partial_x \omega_t(x,t) 
- \tilde{\mathfrak{H}}^{(\Sigma)}(-\imath \partial_x) \omega_t(x,t) 
\label{eq:MasterEquation}
\end{equation}
where the operator $\tilde{\mathfrak{H}}^{(\Sigma)}(-\imath \partial_x)$
is defined by expanding the function  $\tilde{\mathfrak{H}}^{(\Sigma)}(k)$
in a Taylor series around $k=0$ and by replacing $k$ by $-\imath \partial_x$.
The master equation (\ref{eq:MasterEquation}) corresponds to formula (112)
on page 17 in \citet{KleinertPathIntegr}.

In future work we will check if the master equation (\ref{eq:MasterEquation})
includes fractional derivatives.
In particular we will investigate if it can be related to 
distributed order fractional diffusion equations 
\citet{Chechkin,Gorenflo,Sokolov,MeerschaertI,Scheffler} 
that are widely studied due to their connection to ultraslow diffusion.

\section{Conclusions}
We derived the large time limit of a Continuous Time Random Walk 
where the motion of the walker is driven by a stochastic process with
Paretian fluctuations. 
We applied the CTRW model 
to the Heath-Jarrow-Morton model for the term structure 
of interest rates and showed that if the forward rate of interest is driven 
by Paretian fluctuations then the bond price follows a stochastic process 
that behaves like a biased random walk. This means that 
in our model arbitrage opportunities cannot be canceled out by an appropriate choice 
of the drift and the volatility parameters of the HJM model.
Further work will be devoted to deriving asymptotics of distributions
of first hitting times in the CTRW model and to developing 
a theory of pricing options in the Black and Scholes framework.

At the end of writing the manuscript we became aware of a different work
[ \citet{EberleinRaible} ]
devoted to the subject matter. 
We will refer to 
that paper in our future work. 

\section{Appendix}
In this section we derive mathematical results regarding
the small-$s$ expansion (\ref{eq:mathfrakE4}) 
for a convolution of Laplace transforms,
and Laplace transforms (\ref{eq:mathfrakCa}), (\ref{eq:mathfrakCb})
of a power of an inverse logarithm
and of a function
that involves exponentials
and powers of inverse logarithms.

At first we compute the convolution
that is used for computing the Laplace transform of the path integral
(\ref{eq:PathIntegral}). Here:
\begin{eqnarray}
\lefteqn{
\mathfrak{s} := \left( \frac{1}{s (\ln(s))^n} \otimes \frac{1}{s (\ln(s))^m} \right) =
\int_0^s \frac{d \xi}{\xi (\ln(\xi))^m} \frac{1}{(s-\xi) (\ln(s - \xi))^n}}
\label{eq:AppendConvolI} \\
&&= \int_{-\infty}^{\ln(s)} \frac{d z}{z^m} \frac{1}{(s - e^z)(\ln(s - e^z))^n}
= \frac{1}{s (\ln(s))^n} \int_{-\infty}^{\ln(s)} \frac{d z}{z^m} \mathfrak{L}^{(n)}(e^{z - \ln(s)}) 
\label{eq:AppendConvolII} \\
&&= \frac{(-1)^m}{s (\ln(s))^n} \int_{-\ln(s)}^{\infty}
\frac{dw}{w^m} \mathfrak{L}^{(n)}(\frac{e^{-w}}{s})
\label{eq:AppendConvolIII}
\end{eqnarray}
where in (\ref{eq:AppendConvolI})=(\ref{eq:AppendConvolII}) we substituted for $\ln(\xi)$,
then we introduced an auxiliary function
$\mathfrak{L}^{(n)}(x) := (1 - x)^{-1} (1 + \ln(1 - x)/\ln(s))^{-n}$
and finaly in (\ref{eq:AppendConvolII})=(\ref{eq:AppendConvolIII}) we substited for $-z$.
Now we expand the function $\mathfrak{L}^{(n)}$ in a Taylor series 
$\mathfrak{L}^{(n)}(x) = \sum_{p=0}^\infty \mathfrak{l}_p(n) x^p$, 
insert the expansion into (\ref{eq:AppendConvolIII})
and we get:
\begin{eqnarray}
\mathfrak{s} = \frac{(-1)^m}{s(\ln(s))^n}
\sum_{p=1}^\infty \frac{p^{m-1} \mathfrak{l}_p(n) }{s^p} \int_{-p \ln(s)}^\infty w^{-m} \exp(-w) dw
\label{eq:AppendixExpans}
\end{eqnarray}
We recall that the integrals (\ref{eq:AppendixExpans}) are
integral representations of the truncated Gamma function 
\begin{equation}
\Gamma(x,m) := \int_x^\infty w^{-m} \exp(-w) dw
\label{eq:GammaFctDef}
\end{equation}
which 
has a following asymptotic expansion:
\begin{equation}
\Gamma(x,m) = \exp(-x) \sum_{q=0}^\infty (-1)^q \frac{(m+q-1)!}{(m-1)!} \frac{1}{x^{m+q}}
\label{eq:GammaFctExp}
\end{equation}
Inserting the expansion (\ref{eq:GammaFctExp}) into (\ref{eq:AppendixExpans})
we obtain the result (\ref{eq:mathfrakE4}) {\bf q.e.d}.
Note that the coefficients $\mathfrak{l}_p(n)$ in the expansion of the function 
$\mathfrak{L}^{(n)}$ also depend on $s$. They have a form:
$\mathfrak{l}_p(n) = \mathfrak{P}_p^{(n)}\left( 1/\log(s) \right)$
where $\mathfrak{P}_p^{(n)}$ are polynomials of order $p$ and 
$\mathfrak{P}_p^{(n)}=1$.

Now we derive the Laplace transform (\ref{eq:mathfrakCb})
of a function that involves an exponential and powers of the inverse natural logarithm.
We calculate:
\begin{eqnarray}
\lefteqn{
\mathcal{L}_t\left[\frac{e^{-\alpha t}}{(\ln(k t))^n}\right](s) 
 = \frac{1}{k} \left(
\mathcal{I}_{n}^-(\frac{\alpha+s}{k},a) + \mathcal{I}_{n}^+(\frac{\alpha+s}{k},a)
\right)}
\label{eq:LaplaceTransformInvPowLog}
\end{eqnarray}
where 
\begin{equation}
\mathcal{I}_{n}^-(\alpha,a) := \int_0^a \frac{\exp(-\alpha t)}{(\ln(t))^n}dt 
\quad\mbox{and}\quad
\mathcal{I}_{n}^+(\alpha,a) := \int_a^\infty \frac{\exp(-\alpha t)}{(\ln(t))^n}dt 
\end{equation}
In the following we compute the integrals $\mathcal{I}_+^{(n)}(\alpha,a)$ 
and $\mathcal{I}_-^{(n)}(\alpha,a)$. 
For computing the integral $\mathcal{I}_+^{(n)}(\alpha,a)$ we consider
a following auxiliary integral:
\begin{equation}
\mathcal{J}_{n,p}(\alpha,a) := 
\int_a^\infty \frac{\exp(-\alpha t)}{(\ln(t))^n t^p}dt
\quad\mbox{such that}\quad
\mathcal{I}_{n}^+(\alpha,a) = \mathcal{J}_{n,0}(\alpha,a)
\label{eq:AuxilIntDef}
\end{equation}
The integral $\mathcal{J}_{n,p}(\alpha,a)$ differentiated
with respect to the parameter $p$ yields the truncated Gamma function
(\ref{eq:GammaFctDef}). We have
\begin{eqnarray}
\lefteqn{
\frac{d^n \mathcal{J}_{n,p}(\alpha,a)}{d p^n} =
\int_a^\infty e^{-\alpha t} \frac{(-\ln(t))^n}{(\ln(t))^n} \frac{dt}{t^p}} 
\label{eq:AuxilInt1}\\
&&= (-1)^n \alpha^{p-1} \Gamma(\alpha a, p)
  = (-1)^n \frac{e^{-\alpha a} }{\alpha}
    \sum_{q=0}^\infty \frac{(-1)^q}{(\alpha a)^{q}} \frac{(p+q-1)!}{(p-1)!} e^{-p \ln(a)} 
\label{eq:AuxilInt2}
\end{eqnarray}
where on the right hand side in (\ref{eq:AuxilInt1}) we used the identity
$1/t^p = \exp(-p \ln(t))$ and in (\ref{eq:AuxilInt2}) we used the asymptotic expansion
(\ref{eq:GammaFctExp})
of the truncated Gamma function.
Integrating (\ref{eq:AuxilInt2}) $n$ times with respect to $p$ 
from $p$ to infinity and using the fact
(\ref{eq:AuxilIntDef}) that 
$\left. d^m \mathcal{J}_{n,p}(\alpha,a)/ d p^m\right|_{p=\infty} = 0$ 
for $m=0,\ldots,n$
we get:
\begin{equation}
\mathcal{J}_{n,p}(\alpha,a) = 
(-1)^n \frac{e^{-\alpha a} }{\alpha}
\sum_{q=0}^\infty \frac{(-1)^q}{(\alpha a)^{q}} 
F^{(n)}_p\left[ e^{-\xi \ln(a)}\prod_{s=0}^{q-1} (\xi + s)  , \xi\right]
\label{eq:AuxilInt3}
\end{equation}
where 
\begin{equation}
F^{(n)}_p\left[ f(\xi), \xi \right] := 
\int_p^\infty \int_{\xi_1}^\infty \ldots \int_{\xi_{n-1}}^\infty f(\xi_n) d\xi_n \cdot \ldots \cdot d\xi_2 d\xi_1 =
\frac{1}{(n-1)!} \int_p^\infty f(\xi) \xi^{n-1} d\xi
\label{eq:NTimesIntegral}
\end{equation}
We insert the last expression on the right hand side 
in (\ref{eq:NTimesIntegral}) into (\ref{eq:AuxilInt3}),
we evaluate the integral over $\xi$ by successive integration by parts
and we obtain:
\begin{equation}
\mathcal{J}_{n,p}(\alpha,a) = 
\frac{(-1)^n}{(n-1)!} \frac{e^{-\alpha a} }{a^p \alpha}
\sum_{q=0}^\infty \frac{(-1)^q}{(\alpha a)^{q}} 
\sum_{l=0}^{n+q-1} \frac{W_{n-1,q}^{(l)}(p)}{(\ln(a))^{l+1}}
\label{eq:AuxilInt4}
\end{equation}
where
\begin{equation} 
W_{n-1,q}(\xi) := \xi^{n-1} \prod_{s=0}^{q-1} (\xi + s)
\label{eq:DefinitPolynomial}
\end{equation}
and $W_{n-1,q}^{(l)}(p)$ denotes the $l$th derivative evaluted at $\xi = p$.\\
Now we set $p=0$. Since the polynomial $W_{n-1,q}(\xi)$ 
has (\ref{eq:DefinitPolynomial}) a leading power $\xi^{n-1}$
then the derivatives $W_{n-1,q}^{(l)}(0)$ for $l=0,\ldots,n-2$ at zero are equal to zero.
This allows to transform the expression (\ref{eq:AuxilInt4}) for $p=0$
to the following form:
\begin{eqnarray}
\lefteqn{
\mathcal{J}_{n,0}(\alpha,a) = \mathcal{I}_{n}^+(\alpha,a) =} \nonumber \\
&&\frac{(-1)^n}{(n-1)!} \frac{e^{-\alpha a} }{\alpha}
\frac{1}{(\ln(a))^n} 
\sum_{l=0}^\infty \frac{1}{(\ln(a))^l} 
\sum_{q=l}^\infty \frac{W^{(n + l - 1)}_{n + q - 1}(0)}{(\alpha a)^q} (-1)^q
\label{eq:AuxilInt5}
\end{eqnarray}
It is easy to work out the form of the coefficients $W^{(n + l - 1)}_{n + q - 1}(0)$.
They take the form
\begin{equation}
W^{(n + l - 1)}_{n + q - 1}(0) 
= 
\left\{ 
\begin{array}{rr}
(n + l - 1)! (q - 1)! \beta^{(l-1)}_{q - 1} & \mbox{for}\quad l \ge 1 \\
(n - 1)! \delta_{q,0} & \mbox{for}\quad l = 0
\label{eq:Coeffs}
\end{array}
\right.
\end{equation}
where
\begin{equation}
\beta^{(l)}_r := \sum_{1\le \xi_1 < \ldots < \xi_l \le r} (\xi_1 \cdot \ldots \cdot \xi_l)^{-1} \quad\mbox{for $l\ge 1$ and}\quad 
\beta^{(0)}_r = 1
\end{equation}
Inserting (\ref{eq:Coeffs}) into (\ref{eq:AuxilInt5})
and changing the summation index $q=l+r$ to $r$ we get: 
\begin{equation}
\mathcal{I}_{n}^+(\alpha,a) = 
(-1)^n \frac{e^{-\alpha a} }{\alpha}
\frac{1}{(\ln(a))^n} 
\left[ 
\sum_{l=0}^\infty
\left( \begin{array}{c} n + l -1 \\ l \end{array} \right) 
 \frac{\mathcal{A}_l(\alpha a)}{(\ln(a))^l}
\right]
\label{eq:PlusIntegral}
\end{equation}
where
\begin{equation}
\mathcal{A}_l(a) := \sum_{r=0}^\infty 
\frac{l! (l + r - 1)!}{(a)^{l + r}} \beta^{(l-1)}_{l + r - 1} (-1)^{l +  r}
\quad\mbox{and}\quad
\mathcal{A}_0(a) = 1
\label{eq:AsymptoticExp} 
\end{equation}
Now we calculate the integral $\mathcal{I}_{n}^-(\alpha,a)$. The procedure is similar
to that used in the first part of the appendix for deriving result (\ref{eq:mathfrakE4}).
The integral reads:
\begin{eqnarray}
\lefteqn{
\mathcal{I}_{n}^-(\alpha,a) 
= (-1)^n \int_{-\ln(a)}^\infty \exp(-\alpha e^{-z}) \frac{e^{-z}}{z^n} dz =} 
\label{eq:MinusIntegral1}\\
&&(-1)^n \sum_{l=0}^\infty \frac{(-\alpha)^l}{l!} (l+1)^{n-1} \Gamma(-(l+1) \ln(a), n)=
\label{eq:MinusIntegral2}\\
&&(-1)^n \sum_{l=0}^\infty \frac{(-\alpha)^l}{l!} (l+1)^{n-1}
a^{l+1}
\sum_{q=0}^\infty \frac{(n + q - 1)!}{(n - 1)!} \frac{(-1)^q}{(-(l+1)\ln(a))^{n+q}}=
\label{eq:MinusIntegral3}\\
&&\frac{1}{(\ln(a))^n} \sum_{l=0}^\infty \frac{(-\alpha)^l}{(l+1)!} a^{l+1}
\sum_{q=0}^\infty \frac{(n + q - 1)!}{(n - 1)!} \frac{1}{((l+1)\ln(a))^{q}}=
\label{eq:MinusIntegral4}\\
&&\frac{1}{(\ln(a))^n} 
\sum_{l=0}^\infty \frac{(n + l - 1)!}{(n - 1)!} \frac{1}{(\ln(a))^l}
\sum_{q=0}^\infty \frac{(-\alpha)^q}{q!} \frac{a^{q+1}}{(q+1)^{l+1}}
\label{eq:MinusIntegral5}
\end{eqnarray}
where in (\ref{eq:MinusIntegral1}) we substituted for $-\ln(t)$,
in (\ref{eq:MinusIntegral2}) we expanded 
the exponential $\exp(-\alpha e^{-z})$ in a series of $e^{-z}$ and transformed
the resulting integrals to the truncated Gamma function, 
in (\ref{eq:MinusIntegral3}) we made use of the asymptotic expansion 
(\ref{eq:GammaFctExp})
of the truncated Gamma function and finaly in (\ref{eq:MinusIntegral4})
and in (\ref{eq:MinusIntegral5}) we collected together terms with the same
power of the inverse logarithm. 
\begin{equation}
\mathcal{I}_{n}^-(\alpha,a) = 
\frac{1}{\alpha (\ln(a))^n} 
\left[ 
\sum_{l=0}^\infty
\left( \begin{array}{c} n + l -1 \\ l \end{array} \right) 
 \frac{\mathcal{B}_l(\alpha a)}{(\ln(a))^l}
\right]
\label{eq:MinusIntegral}
\end{equation}
where 
\begin{equation}
\mathcal{B}_l(a) = l! \sum_{q=0}^\infty \frac{(-1)^q}{q!} \frac{(a)^{q+1}}{(q+1)^{l+1}}
= \int_0^{a} \left(\ln(\frac{a}{z})\right)^l e^{-z} dz
\label{eq:CoeffIdentity}
\end{equation}
The proof of the last equality in (\ref{eq:CoeffIdentity}) is left to the reader.
Having derived the results (\ref{eq:PlusIntegral}) and (\ref{eq:MinusIntegral})
we complete the computation of the Laplace transform (\ref{eq:LaplaceTransformInvPowLog}).
The result obviously depends on the parameter $a$ that divides the integration domain
into two parts. Since we are interested in the small $s$ limit of the Laplace transform
we choose $a$ to be a decreasing function of $s$, e.g. $a = k/s$.
This conforms to the intuition since $a$ has to be large enough for the asymptotic series
in (\ref{eq:AsymptoticExp}) to converge.
We obtain the following result:
\begin{equation}
\mathcal{L}_t\left[\frac{e^{-\alpha t}}{(\ln(k t))^n}\right](s) 
= \frac{1}{(\alpha + s)} \frac{1}{(\ln(\frac{k}{s}))^n}
\sum_{l=0}^\infty
\left( \begin{array}{c} n + l -1 \\ l \end{array} \right) 
\frac{(\mathcal{B}_l(\theta) + 
(-1)^n e^{-\theta} \mathcal{A}_l(\theta))}
{(\ln(\frac{k}{s}))^p}
\label{eq:LaplTransfFinalRes}
\end{equation}
where $\theta = (\alpha+s)/s$. Therefore the proof of result 
(\ref{eq:mathfrakCb}) is finished.
The Laplace transform (\ref{eq:mathfrakCa}) of a power of an inverse logarithm 
follows immediately by inserting $k=1$ and $\alpha = 0$
into (\ref{eq:LaplTransfFinalRes}). 
The coefficients $\mathcal{C}_p$ and $\mathfrak{k}_p$ 
in (\ref{eq:mathfrakCb}) and (\ref{eq:mathfrakCa})
are expressed via coefficients
$\mathcal{B}_p(\theta)$ and $\mathcal{A}_p(\theta)$ as follows:
\begin{eqnarray}
\mathcal{C}_p &=& (n + p - 1)!/(p! (n - 1)!) \cdot (\mathcal{B}_p(\theta) + 
(-1)^n e^{-\theta} \mathcal{A}_p(\theta))\\
\mathfrak{k}_p &=& (n + p - 1)!/(p! (n - 1)!) \cdot (\mathcal{B}_p(1) + 
(-1)^n e^{-1} \mathcal{A}_p(1))
\end{eqnarray}

\end{document}